\documentclass[11pt]{llncs}
\advance \topmargin -\textheight \divide
\topmargin by 2 \advance \topmargin -1.7in \leftmargin 210mm
\advance \leftmargin -\textwidth \divide \leftmargin by 2 \advance
\leftmargin

\long\def\symbolfootnote[#1]#2{\begingroup%
\def\thefootnote{\fnsymbol{footnote}}\footnote[#1]{#2}\endgroup}

\usepackage[dvips]{graphicx}
\newcommand{\ar}{\rightarrow}

\begin{document}
\title{Modeling the effects of HIV-1 virions and proteins on Fas-induced apoptosis of infected cells}
\titlerunning{Modeling the effects of HIV-1 virions and proteins on Fas-induced apoptosis of infected cells}
\author{John Jack and  Andrei P\u aun$^\dagger$}
 \institute{Department of Computer Science/IfM, Louisiana Tech University,\\ P.O. Box 10348, Ruston, LA 71272
\email{\{johnjack, apaun\}@latech.edu} }
\maketitle
\symbolfootnote[0]{\hspace*{-0.3cm}$\dagger$ \ To whom
correspondence should be addressed. E-mail: apaun@latech.edu}

\begin{abstract}
We report a first in modeling and simulation of the effects of the HIV proteins on the (caspase dependent) apoptotic pathway in infected cells.  This work is novel and is an extension on the recent reports and clarifications on the FAS apoptotic pathway from the literature.  We have gathered most of the reaction rates and initial conditions from the literature, the rest of the constants have been computed by fitting our model to the experimental results reported. Using the model obtained we have then run the simulations for the infected memory T cells, called also latent T cells, which, at the moment, represent the major obstacle to finding a cure for HIV. We can now report that the infected latent T cells have an estimated lifetime of about 42 hours from the moment they are re-activated. As far as we know this is the first result of this type obtained for the infected memory T cells.
\end{abstract}

\section{Introduction}

Apoptosis (or programmed cell death) is the process through which a
multi-cellular organism removes the cells that pose a threat locally
(at the tissue level) or globally to the organism. During the development of organisms,
apoptosis is initiated in order to remove unneeded cells at various developmental stages.
Apoptosis is a form of cellular suicide, an attempt by an organism to cleanly rid itself of
cells that are unwanted, improperly formed, injured, infected, etc. It is a vital process of
multicellular organisms, because of this, the mechanisms of apoptosis are hard-coded in all the cells
of an organism and it can be executed in several ways; we will focus on one particular signaling
pathway, which utilizes members of the Caspase protein family.  The apoptotic pathway has been
recently under a high degree of research focus \cite{FAS,apop2,apop3} due to the fact that it is estimated that more than
50\% of the cancers result due to the inhibition or damage in apoptotic signal transduction.
Using the model from \cite{FAS} we have recently simulated the FAS-mediated apoptosis in
\cite{smitha06} by using a deterministic waiting time algorithm from
\cite{firstDWT} which was then modified and improved into a nondeterministic waiting time algorithm in
\cite{jack07}. The main benefit for using the nondeterministic algorithm proposed by us is the fact that it is a discrete simulation
method (accounting for the discrete nature of the signal transduction pathways) as opposed to differential methods and is also fast as
opposed to the stochastic simulations such as Gillespie's algorithm (see \cite{gillespie76,gillespie}), without losing the nondeterministic behavior.

We are extending in the current paper the model from \cite{jack07} to include HIV-1 proteins in the nondeterministic discrete simulation of the effects of the HIV on the Fas-induced apoptosis. In the following we will give a brief reason for the extension.

The HIV virus is currently causing a global epidemic according to the report published in Geneva on November 21, 2006 by the World
Health Organization. The virus infected roughly 1\% of the global population with about 7.3\% of the infected people dying in 2006.

The HIV-1 virus, which is responsible for 99\% of the HIV infections overall, has some remarkable features. It infects cells from the immune system of the body: mainly CD4+ T cells, but also monocytes and macrophages. The initial infection seems to be through the R5 virus type that uses the co-receptor CCR5 for cell entry as reported in \cite{r5infectsfirst}. Reports have shown that the R5 evolves through mutations into the X4 variant of HIV-1 (using co-receptor CXCR4) in the infected individuals late during the
course of the infection (see for example \cite{r5infectsfirst,x4,x4_2}). There is a significant difference between the R5 and X4 variants of
the virus, R5 is inducing apoptosis in the bystander T cells while the X4 is responsible for the synctium formation
(a multi nucleic cell) which proceeds to the killing of even more immune cells. The fact that the X4 is usually not
present in the first stages of the HIV infection, but is more prevalent in the last stages of AIDS suggests that a normal
immune system could counteract the X4 strand effectively. So one could derive that the immune system is first weakened by the
R5 strand through apoptosis before the X4 strand can ``survive" and collapse it.

In this paper, we focus on simulating the effects on the Fas-induced apoptosis caused by the HIV-1 strand R5 proteins.
We report a simulation for the Fas-mediated pathway in HIV-1 infected cells.  There are no other simulations of this type in existence.
The HIV uses cellular machinery to produce its proteins, in the creation of new virions.  Therefore, when a T cell is activated by the
immune system, it can become infected by HIV, produce HIV proteins (and new virus') for a couple of days,
then die by apoptosis.  Yet, sometimes the CD4+ T cells will become inactive after the HIV genome is incorporated
into the cellular DNA.  This results in a so-called latent infection, which has been of major focus in recent research
(reviewed in \cite{han07}).  In \cite{han07} the authors explain that latently infected cells are persistent in patients
undergoing active HIV treatment (such as HAART).  It is clear that a better understanding of the latent cells will assist
biologists in the development of new treatments.  In Section 2 we will highlight the key aspects of our Fas-induced apoptosis model, Section 3 will
describe the new simulation technique. In Section 4 we review the role of HIV proteins in cell death and in Section 5 we present the results our from model's simulations
and we finish with discussions in Section 6.

\section{Fas-mediated signal cascade}

There are several distinct death receptors, which, when activated, can lead to cellular apoptosis through a tightly regulated molecular
signaling cascade \cite{ash98}.  We have focused our previous efforts on simulating the well-understood Fas-mediated pathway
\cite{jack07}.  According to the literature \cite{rieux03} and \cite{igney02}, understanding the complex signaling cascade of Fas-mediated apoptosis can assist in the creation of remedies for cancer and autoimmune disorders.  In this paper we provide an expansion of our previous model, described in \cite{jack07}, in order to decipher and predict the behavior of HIV-infected cells.
But first, we discuss the Fas-mediated signaling pathway.

There are two versions of the Fas-mediated pathway \cite{scaff98}, they are generally referred to as the type I and type II pathways.  Both pathways begin with the Fas ligand binding to the Fas receptor (CD95) on the cell membrane.  The ligand forces a conformational change of the Fas receptor, which then goes on to recruit the Fas-associated death domain (FADD). Once FADD is bound (up to three FADD per Fasl-Fas site), Procaspase 8 and FLIP can be competitively recruited to the death domain.  FLIP is an inhibitor of the pathway, in that its binding to the death domain removes binding sites for Procaspase 8.  The recruitment of at least two Procaspase 8 molecules can lead to cleavage into active Caspase 8 form.  The divergence in the two pathways occurs after active Caspase 8 is in the cytoplasm; the end result of both pathways is the cleavage of Procaspase 3 into active Caspase 3.  If the initial concentration of Caspase 8 is large enough,
Caspase 3 is directly phosphorylated by Caspase 8 (type I pathway).  Otherwise, Caspase 8 truncates Bid, which binds with two Bax molecules enabling the release of cytochrome c from the mitochondria (type II pathway). Bcl-2 can hinder the release of cytochrome c (thus, inhibiting apoptosis) by binding to Bid, tBid, Bax or the tBid:Bax molecular complexes.  Once released, cytochrome c binds to Apaf and ATP before recruiting and activating
Caspase 9. The active Caspase 9 can then go on to activate Procaspase 3.  XIAP can prohibit apoptosis by binding to Caspase 9, thus preventing the phosphorylation of Caspase 3.

Though the molecules of the Fas-mediated pathways (type I and type II) have been analyzed experimentally (pro- and anti-apoptotic), we believe the only way to understand the dynamics of the complex apoptotic system is through computer simulations.  In \cite{FAS} and \cite{jack07}, three different modeling methods are employed that yield results similar to experimental observations.  We use our ideas from \cite{jack07} and modify the model to incorporate the functions of HIV proteins.

\section{The membrane system model}

There are three schools of thought on modeling molecular interactions.  Microscopic chemistry - that is, the consideration of molecular dynamics for every single molecule in the system (position, momenta of atoms, etc.) - is computationally intractable; the time scale involved in the simulations, uncertainty about cellular components, and the number of atoms involved are factors that make this approach impossible to implement in a computational system.  Differential equations are the macroscopic approach.  This approach has been applied to the Fas pathway previously in \cite{FAS}.

We believe that differential equations are not the best approach for modeling signal cascades involving low numbers of molecules.
We discussed this in general for the Fas-mediated pathway in \cite{jack07} and now we expand our ideas into the realm of HIV
proteins, where proteins like Vpr are at a level of only $\sim$700 molecules upon viral infection \cite{VPRINIT}.  Our approach, referred
to as mesoscopic chemistry, is similar to the microscopic approach.  However, we ignore insignificant molecules, like water and
non-regulated parts of the cellular machinery, and we do not directly model the position and momenta of the molecules, but we model
molecular interactions using statistical equations governing how often and which molecules interact.  Our approach allows finer
modeling of a molecular system than the macroscopic approach, and yet it is more tractable than the microscopic approach.

The membrane system paradigm is based on the compartmental structure found in cells.  Each compartment has different rules and objects, and the system evolves as the rules are carried out based on the objects in the compartments.  As we are modeling molecules in a cell, it is a natural approach to use a membrane system.

We provide now all of the different types of rules (reactions) our membrane system can simulate:

\begin{itemize}
\item[] \hspace*{-0.6cm}{Transformation, complex formation and
dissociation rules:}


$
\begin{array}{cll}
(1) \quad  & [ \; a \; ]_l \ar [ \; b \; ]_l & \\
(2) \quad & [ \; a, \, b \; ]_l \ar [ \; c \; ]_l &\\
(3)  \quad & [ \; a \; ]_l \ar [ \; b, \, c \; ]_l & \\
(4) \quad & [ \; a, \, b, \, c \; ]_l \ar [ \; d \; ]_l &\\
(5)  \quad & [ \; a \; ]_l \ar [ \; b, \, c, \, d \; ]_l & \quad
\mbox{where } a,b,c,d \in \Sigma \mbox{ and } l \in L.
\end{array}
$

These rules are used to specify chemical reactions taking place
inside a compartment of type $l \in L$, more specifically they
represent the transformation of $a$ into $b$; the formation of a
complex $c$ from the interaction of $a$ and $b$; and the
dissociation of a complex $a$ into $b$ and $c$. The rules of type (4)
model the formation of a complex $d$ from the interaction of 3 molecules:
$a$, $b$ and $c$ and the rules of type $5$ model the dissociation of the complex $a$ into 3 molecules $b,\ c,$ and $d$.


\item[] \hspace*{-0.6cm}{Diffusing in and out:}


$
\begin{array}{cll}
(6) \quad  & [ \; a \; ]_l \ar a \; [ \; \; ]_l & \\
(7) \quad  & a \; [ \; \; ]_l \ar [ \; a \; ]_l & \quad \mbox{where
} a \in \Sigma \mbox{ and } l \in L.
\end{array}
$


When chemical substances move or diffuse freely from one compartment
to another we use these types of rules, where $a$ moves from or to a
compartment of type $l$.


\item[] \hspace*{-0.6cm}{Binding and unbinding rules:}


$
\begin{array}{cll}
(8) \quad & a \; [ \; b \; ]_l \ar [ \; c  \; ]_l & \\
(9) \quad & [ \; a \; ]_l \ar b \; [ \; c \; ]_l & \quad \mbox{where
} a,b,c \in \Sigma \mbox{ and } l \in L.
\end{array}
$


Using rules of the first type we can specify reactions consisting in
the binding of a ligand swimming in one compartment to a receptor
placed on the membrane surface of another compartment. The reverse
reaction, unbinding of substance from a receptor, can be described
as well using the second rule.


\item[] \hspace*{-0.6cm}{Recruitment and releasing rules:}


$
\begin{array}{cll}
(10) \quad & \;  a \; [ \; b \; ]_l \ar c \; [ \; \; ]_l & \\
(11) \quad &\; c \; [ \; \; ]_l \ar  a \; [ \; b \; ]_l& \quad
\mbox{where } a,b,c \in \Sigma \mbox{ and } l \in L.
\end{array}
$


With these rules we represent the interaction between two chemicals
in different compartments whereby one of them is recruited from its
compartment by a chemical on the other compartment, and then the new
complex remains in the latter compartment. In a releasing rules a
complex, $c$, located in one compartment can dissociate into $a$ and
$b$, remaining $a$ in the same compartment as $c$, and $b$ being
released into the other compartment.
\end{itemize}

Our interest in the simulator lies in the molecular multiplicity evolution over time.  Our rules are carried out at varying lengths of time, obeying the \emph{Law of Mass Action} - that is, the reaction rate depends proportionally on the multiplicities of the reactants.  Therefore, the number of reactants for a given reaction logically governs the speed at which that reaction will occur.  In our model, the time for each reaction to occur is calculated (Waiting Time or WT), and after sorting all waiting times, the reaction with the least waiting time is applied once (reactants decreased/products increased by one molecule).  The simulation clock is aggregated ($\tau = \tau + $WT), and the WT of each the reaction affected by the applied reaction (i.e., number of reactants of reaction changed) are recalculated.  The process continues until the desired simulation time has been reached.

To calculate the waiting time for each reaction $r$ we have an associated reaction rate constant\footnote{Some of the reaction rate constants are from \cite{FAS} and the HIV protein associate rates are derived from various sources}. For each reaction $r$ we pre-compute a kinetic constant $const_r$: $ const_r = \frac{k_r}{V^{i-1}\times N^{i-1}}$ where $V$ is the volume of the system, $N$ is Avogadro's constant ($6.0221415\times 10^{23}$) and $i$ is the number of reactants involved in the reaction. The time needed for the reaction $r$ to finish will then be computed in the following way: $wt_r=\frac{1}{const_r*|A|*|B|}$ where $|A|$ and $|B|$ represent the multiplicities of the two reactants $A, B$ of the reaction $r$.  On the other hand, if the reaction simulated is a dissociation or other first order reaction involving one reactant, $A$, then the time is computed as $wt_r=\frac{1}{k_r*|A|}$.  Now we shall discuss how the HIV proteins interact with the Fas-mediated pathways (type I and type II), and the method with which we set up a simulation of these effects.

\section{The roles of HIV proteins in apoptosis}
The onset of AIDS is associated with the loss of immune function, due to the death of CD4 T cells (among others).  As part of our simulation, we have focused on the mechanisms through which an infected cell is kept alive in order to produce new provirus', aiding in the spread of the infection.

The method through which an HIV virus infects a cell is well understood.  A spike on the virus (the gp120 molecule) binds to the CD4 receptor of primary T cells, and in conjunction with subsequent binding to a coreceptor (CCR5 or CXCR4), a path is opened for the virus to inject its contents into the cell \cite{chan98} and \cite{wyatt98}.  Reverse transcriptase translates the RNA into DNA and the code is implanted into the cells own DNA for future production, once the cell is activated.  During this time, the immune system fails to detect and destroy the infected cell.  HIV proteins that were packaged in the virion with the RNA interact with various signaling cascades (e.g., Fas pathway) within the cell preparing it for viral reproduction while avoiding annihilation by the immune system.  It is not until the immune system activates the T cell that viral reproduction occurs.

\begin{figure}[h]
\begin{center}
\includegraphics[scale=0.55]{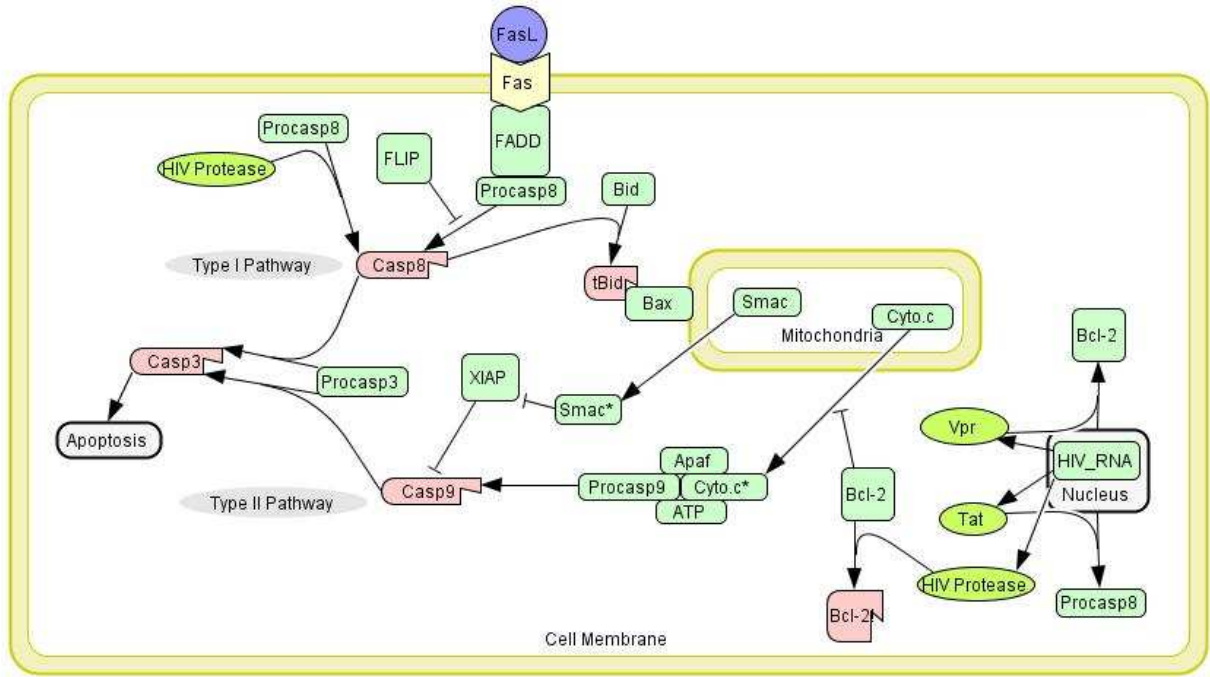}
\end{center}
\caption{The HIV proteins affect the Fas-mediated pathway in various ways.  The picture illustrates the type I and type II pathways.  We consider the activation of Procaspase 3 to be the end of the apoptotic pathway (signifying the death of the cell). The type I pathway involves direct activation of Procaspase
3 by Caspase 8.  The type II pathway requires signal amplification through the release of cytochrome c from the mitochondria resulting in Caspase
9 activating Procaspase 3. Tat upregulates Procaspase 8 and Bcl-2, but it can also down-regulate Bcl-2.  Vpr upregulates Bcl-2 and downregulates Bax.  HIV protease can cleave Bcl-2 into an inactive form and it can also cleave Procaspase
8 into active Caspase 8.}
\label{fig1}
\end{figure}

In this study, we focus on the specific functions of the HIV proteins related to the apoptosis of CD4 T cells; reviews are found in \cite{sellia01}, \cite{ross01} and \cite{alimon03}.  The functions of some of these proteins are still up for debate, however, we have pooled the collective knowledge of the biological community in order to categorize and model the described functions of various HIV proteins.   For an illustration of the FAS-mediated apoptosis pathway and the involvement of the HIV proteins, we refer the reader to Fig. \ref{fig1}.

The HIV protein, Tat, has shown both pro-apoptotic and anti-apoptotic behavior.  In \cite{bartz99}, the authors were able to show that Tat expression in the cell leads to up-regulation of Procaspase
8 (proapoptotic molecule).  Tat has been associated with upregulation of Bcl-2 (antiapoptotic molecule) \cite{fink95}.  Conversely, \cite{sellia01} argues that Tat downregulates Bcl-2, resulting in increased sensitivity to the Fas-mediated type II pathway.  Tat has also been implicated in the upregulation of the Fas ligand in infected cells \cite{bartz99} and \cite{yang02}.  However, the upregulation is assumed to be a mechnism to destroy CTLs (thus, keeping the cell alive) and bystander cells through Fas-mediated apoptosis, the increase of Fas ligand does not appear to affect apoptosis of the infected cell \cite{sellia01}.

Another HIV protein, Vpr, has been implicated in both prohibiting and assisting apoptosis.  Since Vpr is located in the virions, a low concentration is avaible early in the infected cell life-cycle (assumed to be ~700 molecules according to \cite{VPRINIT}).  At low multiplicity, Vpr has been shown to prohibit apoptosis by up-regulating Bcl-2 and down-regulating Bax (proapoptotic molecule) \cite{conti98}.

HIV protease, which is responsible for cleaving the proteins created from the viral genetic code, has also been shown to affect the signaling cascade in apoptosis.  The HIV protease has been shown to cleave Bcl-2 into a deactive state \cite{strack96}, while it is also responsible for cleaving Procaspase 8 into active Caspase 8 form \cite{nie02}, both events lead to an increased susceptibility for Fas-mediated apoptosis.


The HIV RNA travels to the nucleus and becomes incorporated into the cellular genetic code.
While the T cell is activated by the immune
system, the HIV proteins are manufactured through the cellular machinery (transcription into HIV RNA and translation through the ribosomes
in the cytoplasm).  The simulator makes stochastic choices on where to send the
Tat, Vpr, and HIV Protease (cytoplasm, mitochondria, or extracellular).  Using the information on the three HIV proteins, we were able
to create 11 new rules for our simulator.  Our results are discussed below.

\section{Results of the Infected Model}

Based on the rules in Appendix A we simulated two different version of T cells.  What we refer to as the ''Non-latent" is a simulation of the T cell just after the contents of a virion have infected it, so there is Vpr active in the cytoplasm and the RNA is on its way to being incorporated into the DNA of the active CD4 cell.  We call the second simulation method ''Post-latent", which is a simulation of the reactivation of an infected, but latent CD4 cell.  We seek to achieve greater understanding on the nature of latently infected cells.

\begin{figure}[h]
\begin{center}
\includegraphics[scale=0.3]{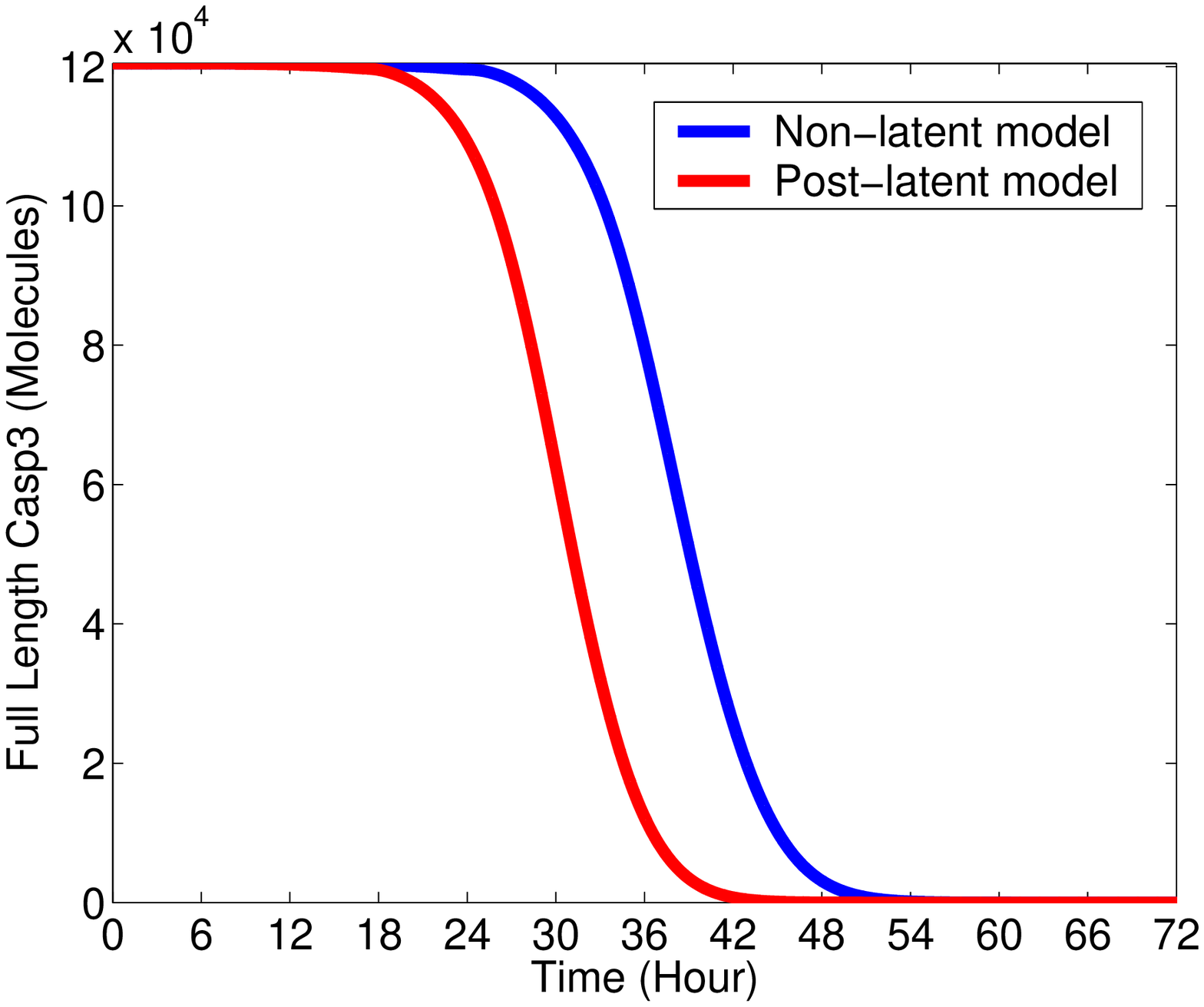}
\includegraphics[scale=0.3]{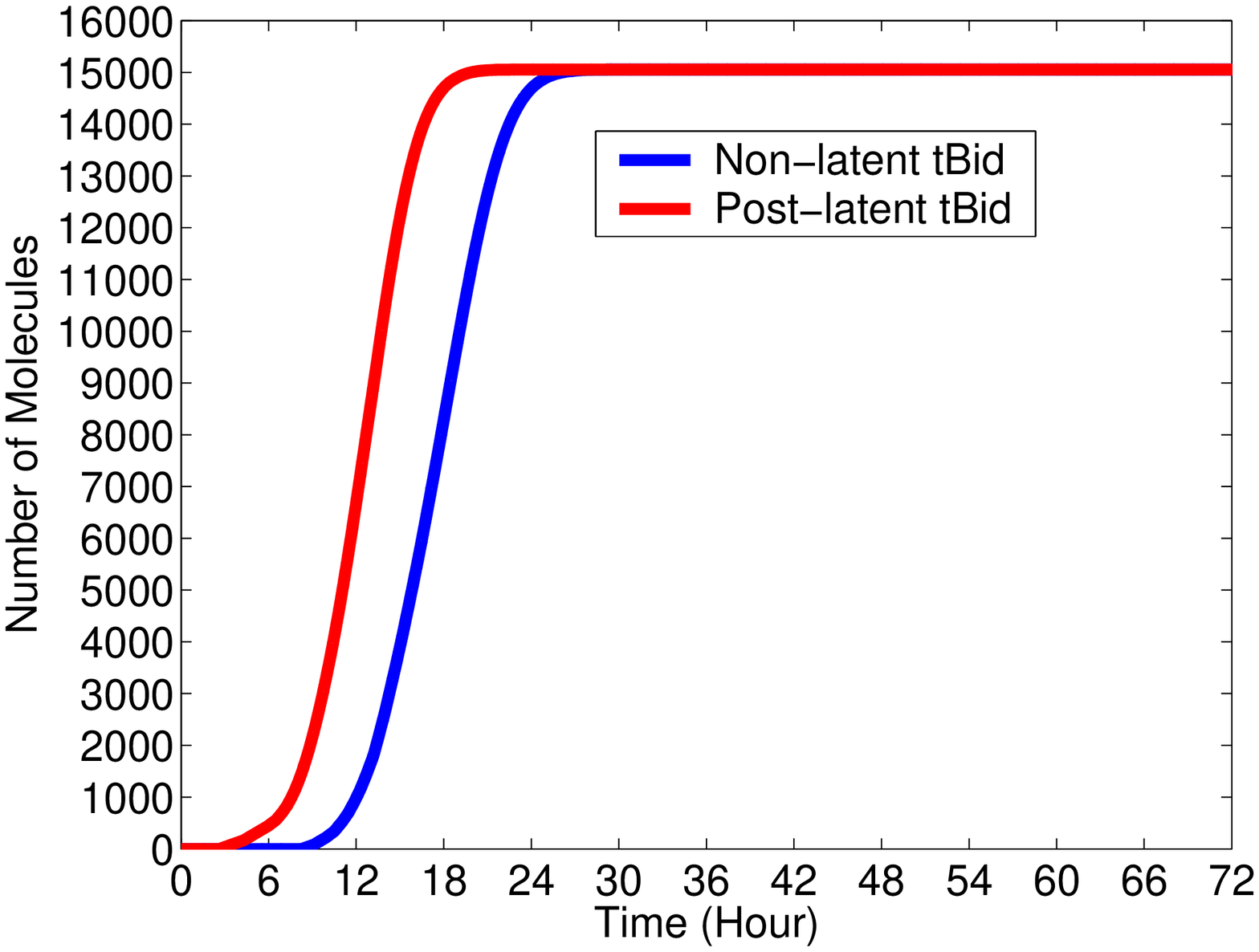}
\end{center}
\caption{On the left, we see cell death for the latently infected cell occuring in ~42 hours and freshly infected cell dying in ~49 hours.  On the right, we illustrate the rates of truncations for Bid, a signal for the induction of the type II pathway.}
\label{fig2}
\end{figure}

As we consider activation of Procaspase 3 to be an indication of the onset of apoptosis, we see in Fig. \ref{fig2} that both the Post-latent and Non-latent cells die in approximately two days.  Our simulations do indicate that reactivated latent cells will die more quickly than freshly infected cells.  The lack of Vpr molecules in the latent cells is most likely the cause, as the anti-apoptotic effect of the protein would not be available.  Also in Fig. \ref{fig2} we show the truncation of Bid, a necessary step in the induction of the type II pathway.  Active
Caspase 8 is responsible for the truncation of Bid, so we again see the affects of the increased rate of Caspase 8 activation.  The amplification of the signal through the mitochondria occurs earlier in the Post-latent than in the Non-latent simulation.

\begin{figure}[h]
\begin{center}
\includegraphics[scale=0.3]{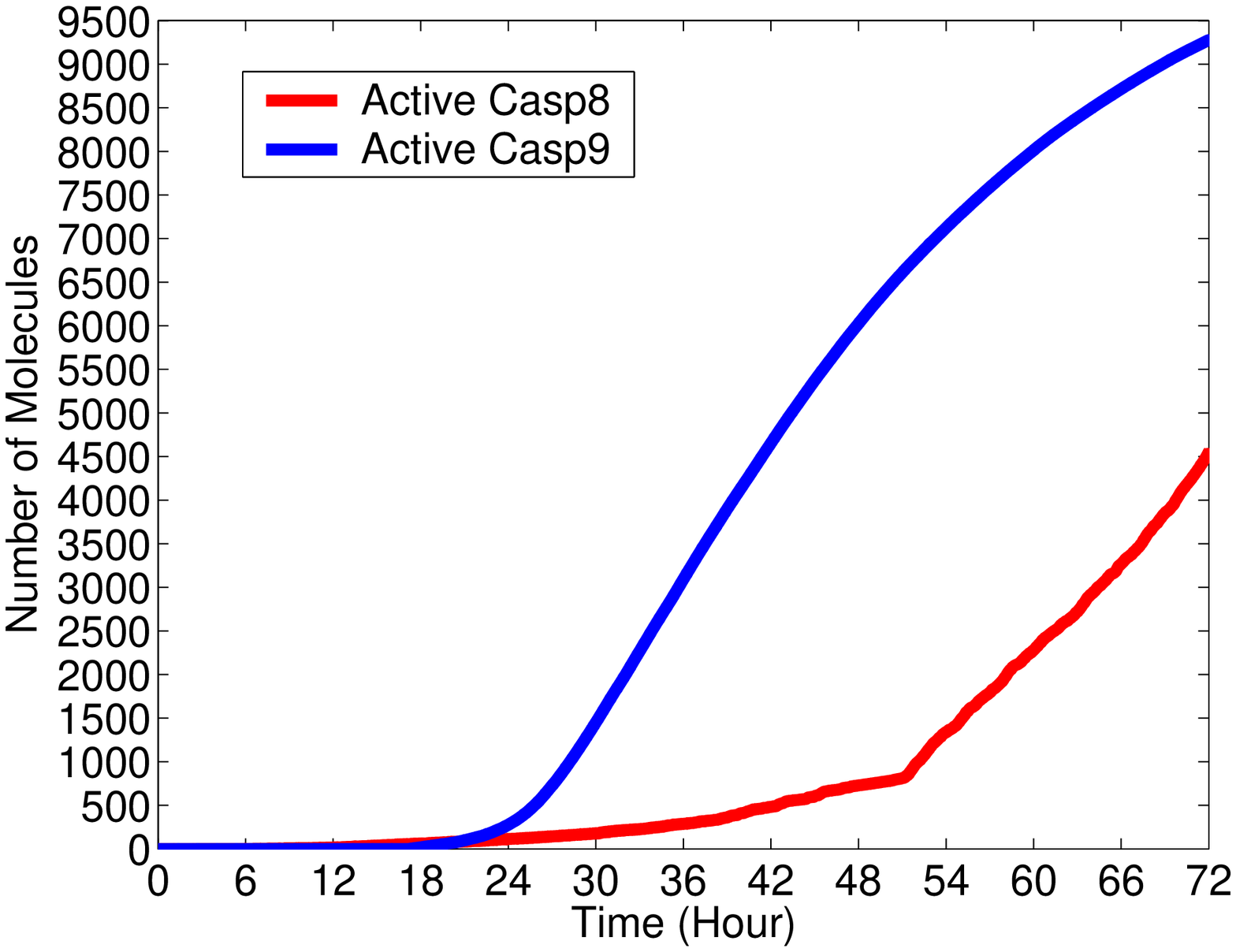}
\includegraphics[scale=0.3]{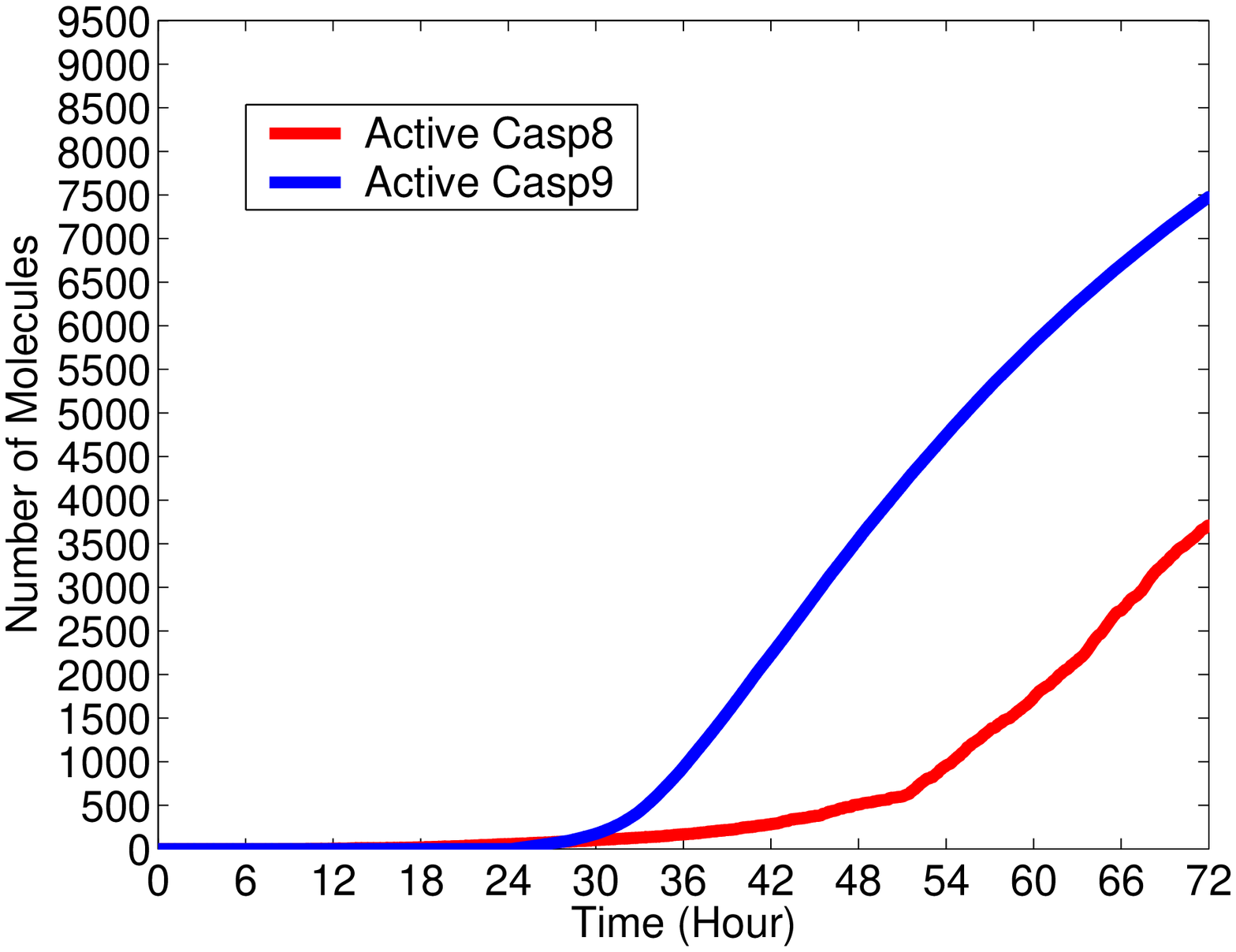}
\end{center}
\caption{These graphs show the increase of active Caspase 8 and Caspase 9 molecules over the entire simulation run.  On the left is the Post-latent simulation and on the right is the Non-latent run.}
\label{fig3}
\end{figure}

In an HIV infected cell, Procaspase 8 can be activated by two ways, interactions with FADD or HIV Protease.  The upregulation of Procaspase 8 by Tat increases the activation rate of Procaspase 8 (because it increases the chance of Procaspase 8 interacting with the existing FADD or HIV Protease molecules).  As the number of molecules of Fas ligand is the same between both simulations, the increased rate of apoptosis is due to the creation and activity of HIV Protease and Tat.  In Fig. \ref{fig3}, we show the activation of Procaspase 8
 and Procaspase 9 for each of the simulations.

\begin{figure}[h]
\begin{center}
\includegraphics[scale=0.3]{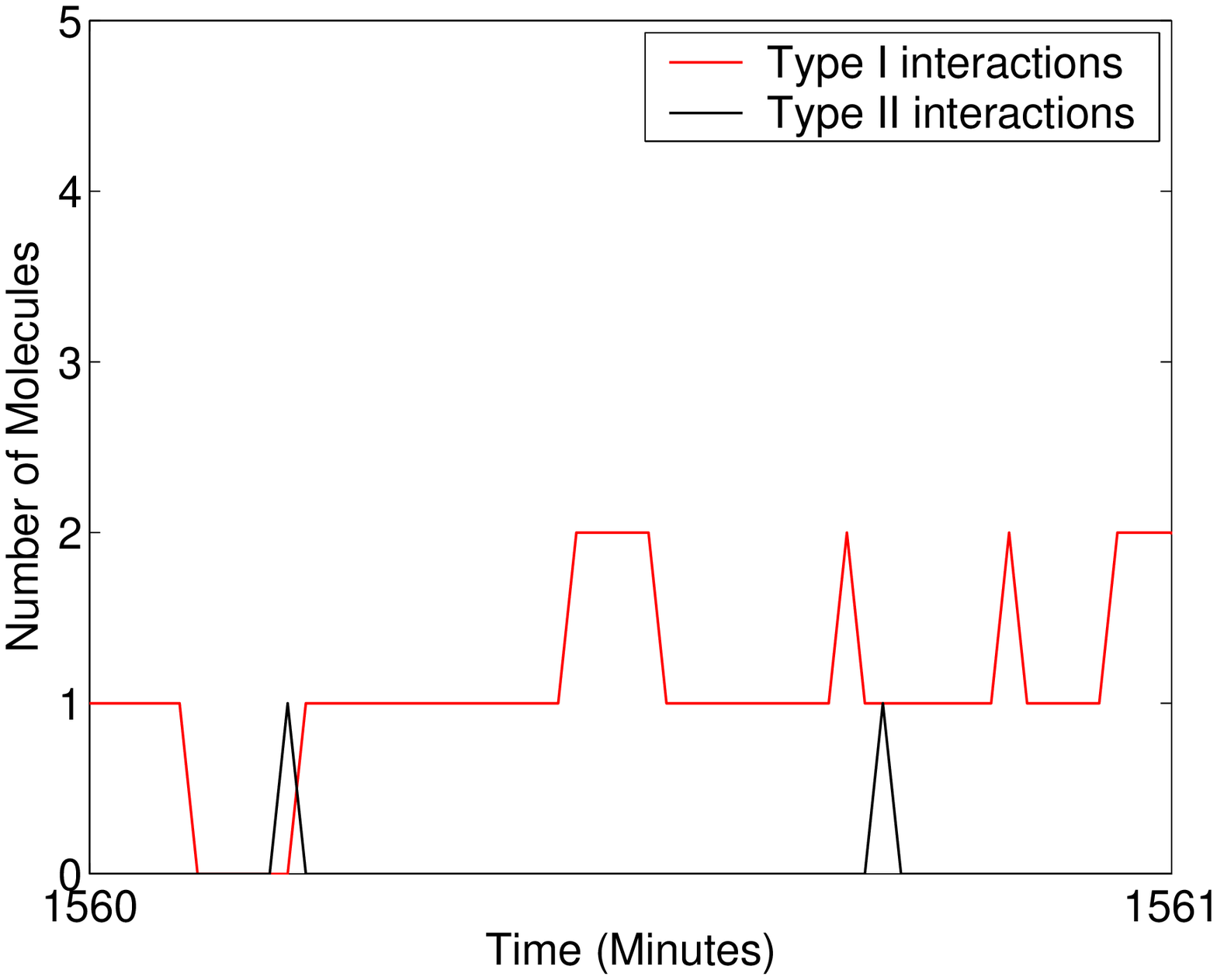}
\includegraphics[scale=0.3]{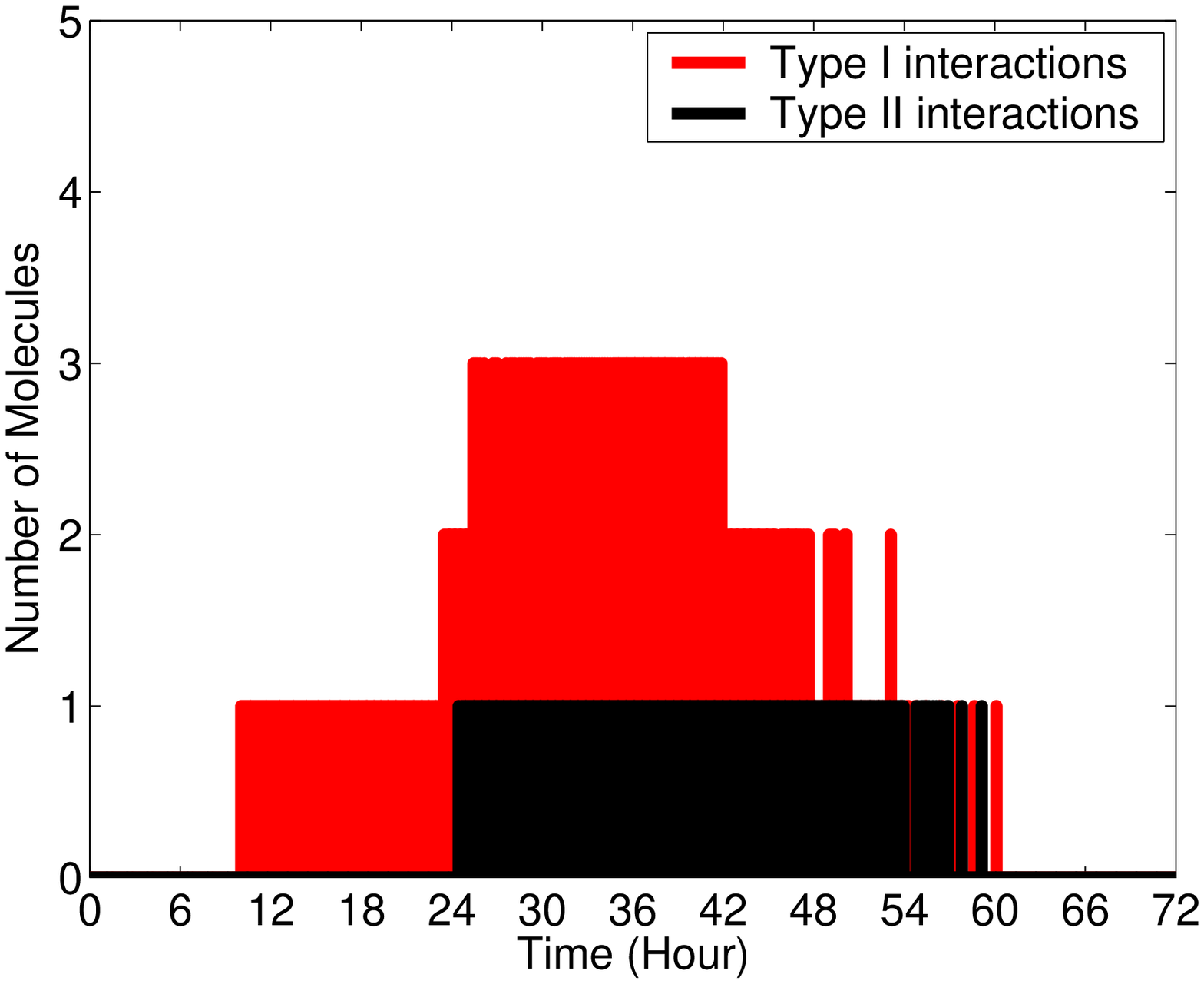}
\end{center}
\caption{These graphs show the type I and type II pathways working together to activate Procaspase 3 during the non-latent simulation. The Type I interactions are active Caspase 8 binding with Procaspase 3, and the Type II interactions are Caspase 9 binding with Procaspase 3. On the right we see the overall picture for the whole three days of simulation.  On the left we can see a window of one whole minute from the simulation (from 26 hours to 26 hours and 1 minute).}
\label{fig5}
\end{figure}

We next seek to determine whether the Procaspase 3 is being activated through the type I or the type II pathway.  According to our rule set, an interaction between Procaspase 3 and active Caspase 8 or Caspase 9 can have two outcomes, resulting in activation of the Procaspase 3 molecule or not.  Interestingly, it appears that the two pathways work together to annihilate the cell.  The Non-latent simulation shows the first interaction of a molecule of Procaspase 3 occurs with a Caspase
8 molecule just after 10 hours into the run.  It isn't until a 14 hours later (24.5 hours into the run, allowing for signal transduction through the mitochondria) that we begin to see Procaspase 3 interactions with active Caspase 9.  As we have shown in \cite{jack07}, given a sufficiently high concentration of procaspase 8 in the T cell, there is no need for signal amplification.  However, the initial level of Procaspase 8 in our model reflects Jurkat T Cells \cite{FAS}, and we are seeing cooperation between Caspase 8 and Caspase
9 leading to cellular apoptosis.  In Fig. \ref{fig5} we get an overall image of the effects of both pathways on Procaspase 3 activation.

\begin{figure}[h]
\begin{center}
\includegraphics[scale=0.3]{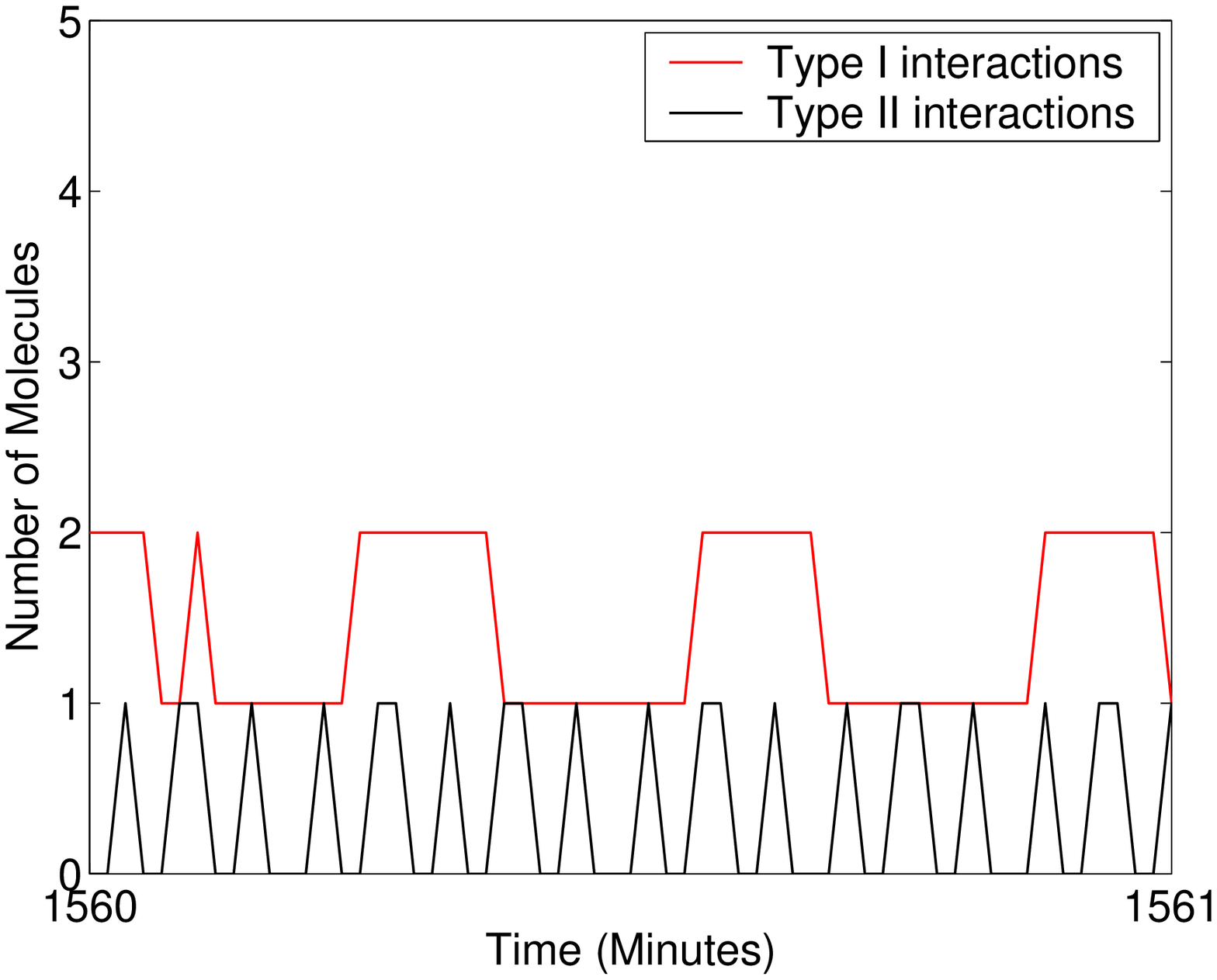}
\includegraphics[scale=0.3]{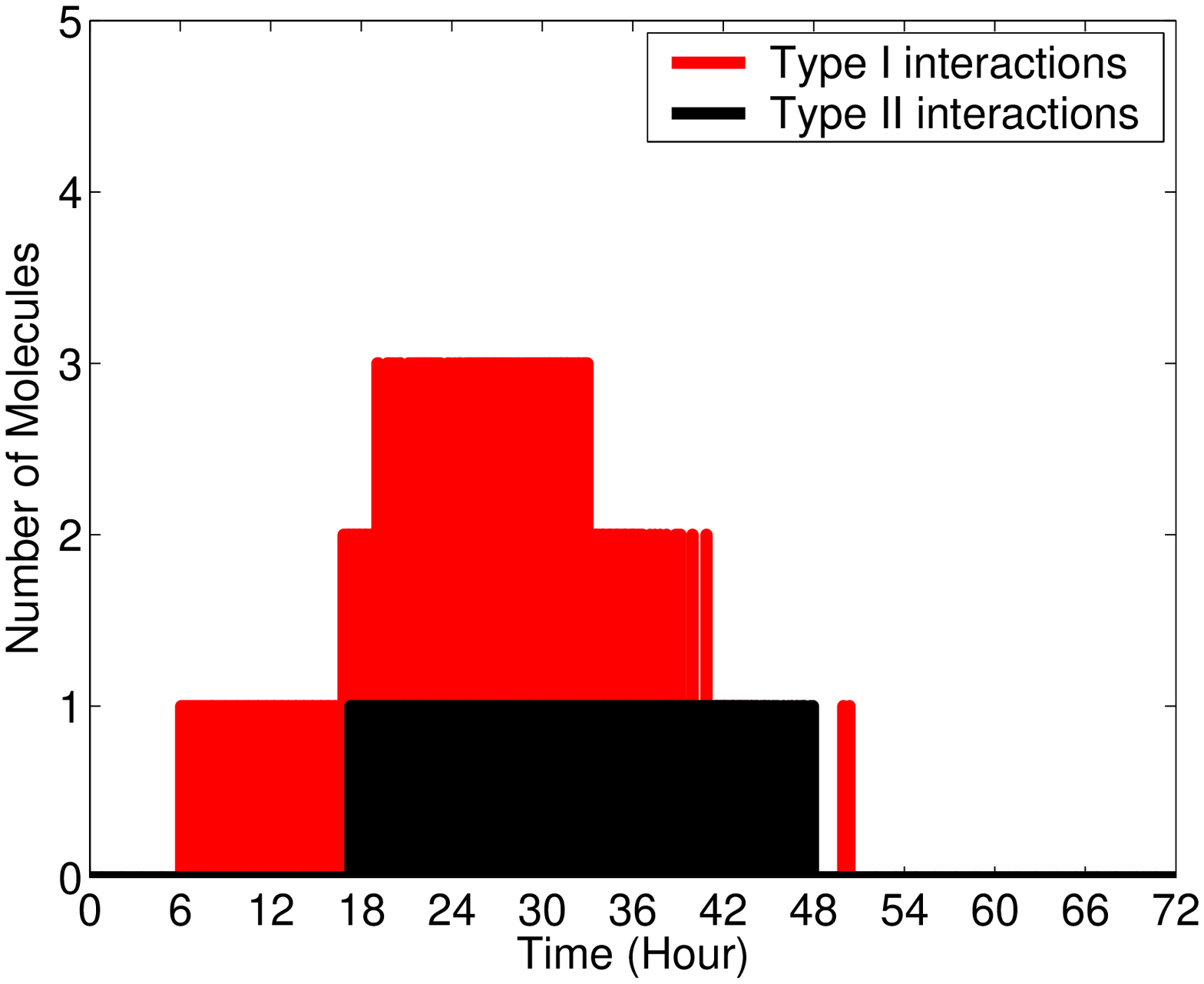}
\end{center}
\caption{These graphs show the type I and type II pathways working together to activate Procaspase 3 during the post-latent simulation. The Type I interactions are active Caspase 8 binding with Procaspase 3, and the Type II interactions are Caspase 9 binding with Procaspase 3. On the right we see the overall picture for the whole three days of simulation.  On the left we can see a window of one whole minute from the simulation (from 26 hours to 26 hours and 1 minute).}
\label{fig8}
\end{figure}

We find similarities in the Procaspase 3 activation between the Non-latent and Post-latent simulation runs.  In the Post-latent run, we see type I interactions occur about 6 hours into the simulation, while type II molecular binding occurs after 17 hours (11 hours after the type I interactions).  Fig. \ref{fig8} show us an the overall interactions between Procaspase 3 with active Caspase
8 or Caspase 9.

The graphs in Fig. \ref{fig5} and \ref{fig8} imply that there are more interactions in the type I pathway.  This does not necessarily mean that the type I pathway has a greater responsibility in activation of Caspase 3 than the type II pathway.  Caspase 3 activation is a two step process.  First, Caspase 8 or 9 binds to the Procaspase 3 molecule, then after a period of time the binding ends with either an active Caspase 3 or the original Procaspase 3 molecule.  Technically, based on the kinetic rates and the concentration of molecules involved, Caspase 8 can stay binded to Procaspase 3 for longer than Caspase 9.  Therefore, though it appears that Caspase 8 binds to Procaspase 3 more frequently, it might not be the case that this results in more activation of Caspase 3 via Caspase 8 than Caspase 9.  The graphs illustrate this possibility by the fact that you see constant jumps for the type II interactions (Caspase 9 and Procaspase 3), which implies less waiting time during the binding period.



\section{Discussion}
Based on the biological evidence from the literature, we were able to construct a simulation of the effects of HIV proteins on the Fas-mediated apoptosis pathway.  We feel that our unique approach provides insight into the molecular dynamics of the Fas pathway.  Our simulator leads to apoptosis in approximately two days, which is in agreement with observed results.  But, we have shown that cooperation between the type I and type II pathway seems to be the method of the cell death.

In this preliminary simulator, we have focused primarily on the difference between reactivated latent cells and actively infected cells, but we are looking to refine and expand our technique.  We will apply new information on the nature of latent HIV infected cells as it becomes available.  However we will also explore alternative HIV models.  For instance, it would be interesting to model the effects HIV proteins on bystander cell apoptosis.  The type 1 HIV has been implicated in reducing T cell levels by not only destroying the cell it invades, but through the killing of so-called bystander (uninfected) cells \cite{fink95}.  Various mechanisms have been debated as to the means through which the virus causes the destruction of the bystander (uninfected) cells.  Along with Fas-induced apoptosis, other possible mechanisms for bystander cell death are reviewed in \cite{sellia01}, \cite{ross01} and \cite{alimon03}. Upon being exocytised by an infected cell, several distinct types of HIV proteins can affect neighboring bystander cells.

Soluble and membrane-bound Env can bind to the CD4 receptor of bystander cells (as is the case during infection).
Along with subsequent binding to a co-receptor, it can induce apoptosis.  In \cite{cica00} and \cite{biard00},
it has been shown that ligation of the CD4 receptor by Env, is sufficient to increase apoptosis in bystander cells. The reasons for
increased apoptosis following Env-CD4 binding can be attributed to Bcl-2 down-regulation \cite{hash97}, increased
Caspase 8 activation \cite{alge02}, and upregulation of Fas \cite{oyaizu94}, FasL and Bax \cite{sellia01}.

Extracellular Tat can be endocytised by a bystander cell, resulting in pro-apoptotic behavior.  The addition of Tat to a culture of uninfected cells has been shown to increase apoptosis \cite{mcclosk97}.  Endocytised Tat can up-regulate levels of
Caspase 8 \cite{bartz99} and increase expression of the Fas ligand \cite{sellia01}, similar to its effects in infected cells.
Extracellular Vpr can disrupt the mitochondrial membrane, leading to increased translocation of cytochrome c \cite{sellia01}.

Lastly we would like to note the fact that the experimental information about the latent HIV infected T cells is scarce, the main reason being
the fact that these cells are found in small numbers in vivo, thus it is rather hard to obtain sufficient numbers for experiments in vitro.
We hope that our results will help guide the researchers in the lab and we are looking forward to experimental results about these enigmatic cells
that will suggest improvements and refinements of our model.

\newpage

\appendix

\section{The Model}

Our model consist of the following P system:

$$
\Pi = ( \Sigma, L, \mu, M_1, M_2, M_3, M_4, M_5, R_1, R_2, R_3, R_4, R_5)
$$

where:

\begin{itemize}

\item In the alphabet we represent the proteins and complexes of
proteins involved in the signalling transduction:

$\Sigma = \{ $HIVRNA, Tat, Vpr, HIV Protease, FASL, FAS, FASC, FADD, FASC-FADD, FASC-FADD$_2$,
FASC-FADD$_3$, FASC-FADD$_2$-CASP8, FASC-FADD$_3$-CASP8,
FASC-FADD$_2$-FLIP, FASC-FADD$_3$-FLIP, FASC-FADD$_2$-CASP8$_2$,
FASC-FADD$_3$-CASP8$_2$, FASC-FADD$_2$-CASP8-FLIP,
FASC-FADD$_3$-CASP8-FLIP, FASC-FADD$_2$-FLIP$_2$,
FASC-FADD$_3$-FLIP$_2$, FASC-FADD-CASP8, FASC-FADD-FLIP,
CASP8, FLIP, FASC-FADD$_3$-CASP8$_3$, FASC-FADD$_3$-CASP8$_2$-FLIP,
FASC-FADD$_3$-CASP8-FLIP$_2$, FASC-FADD$_3$-FLIP$_3$,
CASP8$_2^{P41}$, CASP8$_2^*$, CASP3, CASP$8_2^*$-CASP3, CASP$3^*$,
\linebreak CASP8$_2^*$-Bid, tBid, Bid, Bax, tBid-Bax, tBid-Bax$_2$,
Smac, Smac$^*$, Cyto.c, Cyto.c$^*$, XIAP, Smac$^*$-XIAP, Apaf,
Cyto.c$^*$-Apaf-ATP, CASP9, Cyto.c$^*$-Apaf-ATP-CASP9,
Cyto.c$^*$-Apaf-ATP-CASP9$_2$, CASP9$^*$, CASP9$^*$-CASP3,
CASP9-XIAP, CASP3$^*$-XIAP, Bcl$2$, Bcl$2$-Bax, Bcl$2$-tBid$\}.$\footnote{Smac is
an abbreviation for second mithocondria-derived activator of
caspase.}

\item The set of labels, $L = \{ e,s,c,m,n \}$, can also be used to
identify the different compartments where the signalling
transduction takes place, namely, the environment ($e$), the cell
surface ($s$), the cytoplasm ($c$), the mitochondria ($m$), and the nucleus ($n$).

\item In the membrane structure $\mu=[_1[_2[_3[_4\ ]_4 [_5 \ ]_5 ]_3 ]_2 ]_1$, the
compartments are represented. The compartments 1, 2, 3, and 4 can be
labeled respectively with $e$, $s$, $c$, $m$, and $n$.

\item $M_i = \{ w_i, l_i\} $ where the multiset $w_i$ represents
the initial amount of molecules
in the compartment and $l_i$ is the label associated
with the compartment. For Example:

\begin{itemize}

\item $l_1 = e$ and $M_1 = \{ $FASL$^{301} \}$

\item $l_2 = s$ and $M_2 = \{ $FAS$^{602}\}$

\item $l_3 = c$ and $M_3 = \{ $FADD$^{10038}$, CASP8$^{20071}$,
FLIP$^{48779}$, CASP3$^{120442}$, Bid$^{15055}$, \\
\hspace*{2.8cm} Bax$^{50182}$, XIAP$^{18066}$, Apaf$^{60221}$, ATP$^{6022141}$
CASP9$^{12044}$, Vpr$^{700} \}$

\item $l_4 = m$ and $M_4 = \{$ Smac$^{60221}$, Cyto.c$^{60221}$ Bcl$2^{451661} \}$

\item $l_5 = n$ and $M_5 = \{$ HIVRNA$^{2} \}$

\end{itemize}

\item $R_1,R_2, R_3$, $R_4$, and $R_5$ are the sets of rules associated with
the environment, cell surface, cytoplasm, mitochondria, and the nucleus. These
rules represent the chemical reactions that take place in each
compartment or region of the cell.

\begin{itemize}

\item $R_1 = \{ r_1 \}$

\item $R_2 = \{ r_2, r_4, r_6, r_8, r_{10}, r_{11}, r_{12},
r_{14}, r_{16}, r_{17}, r_{18}, r_{20}, r_{22}, r_{24}, r_{26},
r_{28}, r_{30}, r_{32}, r_{34}, \\ \hspace*{1.1cm}r_{36}, r_{38},
r_{40}, r_{42}, r_{44}, r_{46}, r_{48}, r_{50}, r_{52}, r_{54},
r_{56}, r_{58}, r_{60}, r_{62}, r_{63}, r_{64}, r_{65}, r_{66} \}$

\item $R_3 = \{ r_3, r_5, r_{7}, r_{9}, r_{11}, r_{13}, r_{15},
r_{17}, r_{19}, r_{21}, r_{23}, r_{25}, r_{27}, r_{29}, r_{31},
r_{33}, r_{35}, r_{37}, r_{39}, \\\ \hspace*{1.1cm}r_{41}, r_{43},
r_{45}, r_{47}, r_{49}, r_{51}, r_{53}, r_{55}, r_{57}, r_{59},
r_{61}, r_{67}, r_{68}, r_{69}, r_{70}, r_{71}, r_{72}, r_{73}, \\
\hspace*{1.1cm} r_{74}, r_{75}, r_{76}, r_{77}, r_{78}, r_{79},
r_{80}, r_{81}, r_{82}, r_{83}, r_{84}, r_{85}, r_{86}, r_{87},
r_{88}, r_{89}, r_{90}, r_{91}, \\ \hspace*{1.1cm} r_{92}, r_{93},
r_{94}, r_{95}, r_{96} r_98, r_99, r_{100}, r_{101}, r_{102}, r_{103}, r_{104}, r_{105}, r_{106}\}$

\item $R_4 = \{ r_{97} \}$

\item $R_5 = \{ r_{107}, r_{108}, r_{109}, r_{110} \}$

\end{itemize}
\end{itemize}

We list the four sets of rules we simulated, which are based upon the apoptosis pathway.
Each set incorporates a mechanism for Bcl2 involvement in the inhibition of apoptosis.\medskip

\noindent$
\begin{array}{clc}
\mbox{label} & \ \ \ \ \mbox{ rule }& \mbox{rate}\\
r_1:& FASL  [ \;  FAS  \; ]_s \ar   [ \;  FASC  \;]_s&k_{1f}\\
r_2:& [ \;   FASC   \; ]_s \ar   FASL   [ \;   FASC   \;
]_s&k_{1r}\\
r_3:&FASC   [ \;   FADD    \; ]_c \ar   FASC:FADD  [ \;
  \; ]_c&k_{2f}\\

r_4:&FASC:FADD   [ \;      \; ]_c \ar   FASC   [ \;
FADD   \; ]_c&k_{2r}\\

r_5:&FASC:FADD   [ \;   FADD   \; ]_c \ar   FASC:FADD _2
  [ \;     \; ]_c&k_{2f}\\

r_6:&FASC:FADD _2    [ \;     \; ]_c \ar   FASC:FADD   [
\;   FADD   \; ]_c&k_{2r}\\

r_7:&FASC:FADD _2    [ \;   FADD   \; ]_c \ar
FASC:FADD _3    [ \;     \; ]_c&k_{2f}\\

r_8:&FASC:FADD _3    [ \;     \; ]_c \ar   FASC:FADD _2
[ \;   FADD   \; ]_c&k_{2r}\\

r_9:&FASC:FADD _2 :CASP8   [ \;   FADD   \; ]_c \ar
FASC:FADD _3 :CASP8   [ \;     \; ]_c&k_{2f}\\

r_{10}:&FASC:FADD _3 :CASP8   [ \;     \; ]_c \ar
FASC:FADD _2 :CASP8   [ \;   FADD   \; ]_c&k_{2r}\\

r_{11}:&FASC:FADD _2 :FLIP   [ \;   FADD   \; ]_c \ar
FASC:FADD _3 :FLIP   [ \;     \; ]_c&k_{2f}\\

r_{12}:&FASC:FADD _3 :FLIP   [ \;     \; ]_c \ar
FASC:FADD _2 :FLIP   [ \;   FADD   \; ]_c&k_{2r}\\

r_{13}:&FASC:FADD _2 :CASP8 _2    [ \;   FADD   \; ]_c
\ar   FASC:FADD _3 :CASP8 _2    [ \;     \; ]_c&k_{2f}\\

r_{14}:&FASC:FADD _3 :CASP8 _2    [ \;     \; ]_c \ar
FASC:FADD _2 :CASP8 _2    [ \;   FADD   \; ]_c&k_{2r}\\

r_{15}:&FASC:FADD _2 :CASP8:FLIP   [ \;   FADD    \; ]_c
\ar   FASC:FADD _3 :CASP8:FLIP   [ \;     \; ]_c&k_{2f}\\

r_{16}:&FASC:FADD _3 :CASP8:FLIP   [ \;     \; ]_c \ar
FASC:FADD _2 :CASP8:FLIP   [ \;   FADD   \; ]_c&k_{2r}\\

r_{17}:&FASC:FADD _2 :FLIP _2    [ \;   FADD   \; ]_c \ar
  FASC:FADD _3 :FLIP _2    [ \;     \; ]_c&k_{2f}\\

r_{18}:&FASC:FADD _3 :FLIP _2    [ \;     \; ]_c \ar
FASC:FADD _2 :FLIP _2    [ \;   FADD   \; ]_c&k_{2r}\\

r_{19}:&FASC:FADD:CASP8   [ \;   FADD   \; ]_c \ar
FASC:FADD _2 :CASP8   [ \;     \; ]_c&k_{2f}\\

r_{20}:&FASC:FADD _2 :CASP8   [ \;     \; ]_c \ar
FASC:FADD:CASP8   [ \;   FADD   \; ]_c&k_{2r}\\

r_{21}:&FASC:FADD:FLIP   [ \;   FADD   \; ]_c \ar
FASC:FADD _2 :FLIP   [ \;     \; ]_c&k_{2f}\\

r_{22}:&FASC:FADD _2 :FLIP   [ \;     \; ]_c \ar
FASC:FADD:FLIP   [ \;   FADD   \; ]_c&k_{2r}\\

r_{23}:&FASC:FADD _3    [ \;   CASP8   \; ]_c \ar
FASC:FADD _3 :CASP8   [ \;     \; ]_c&k_{2f}\\

r_{24}:&FASC:FADD _3 :CASP8   [ \;     \; ]_c \ar
FASC:FADD _3    [ \;   CASP8   \; ]_c&k_{2r}\\

r_{25}:&FASC:FADD _3    [ \;   FLIP   \; ]_c \ar
FASC:FADD _3 :FLIP   [ \;     \; ]_c&k_{3f}\\

r_{26}:&FASC:FADD _3 :FLIP   [ \;     \; ]_c \ar
FASC:FADD _3    [ \;   FLIP   \; ]_c&k_{3r}\\

r_{27}:&FASC:FADD _3 :CASP8    [ \;   CASP8   \; ]_c \ar
  FASC:FADD _3 :CASP8 _2    [ \;     \; ]_c&k_{3f}\\

r_{28}:&FASC:FADD _3 :CASP8 _2    [ \;     \; ]_c \ar
FASC:FADD _3 :CASP8   [ \;   CASP8   \; ]_c&k_{3r}\\

r_{29}:&FASC:FADD _3 :CASP8   [ \;   FLIP   \; ]_c \ar
FASC:FADD _3 :CASP8:FLIP   [ \;     \; ]_c&k_{3f}\\

r_{30}:&FASC:FADD _3 :CASP8:FLIP   [ \;     \; ]_c \ar
FASC:FADD _3 :CASP8   [ \;   FLIP   \; ]_c&k_{3r}\\

r_{31}:&FASC:FADD _3 :FLIP   [ \;   CASP8   \; ]_c \ar
FASC:FADD _3 :CASP8:FLIP   [ \;     \; ]_c&k_{3f}\\

r_{32}:&FASC:FADD _3 :CASP8:FLIP   [ \;     \; ]_c \ar
FASC:FADD _3 :FLIP   [ \;   CASP8   \; ]_c&k_{3r}\\

r_{33}:&FASC:FADD _3 :FLIP   [ \;   FLIP   \; ]_c \ar
FASC:FADD _3 :FLIP _2    [ \;     \; ]_c&k_{3f}\\

r_{34}:&FASC:FADD _3 :FLIP _2    [ \;     \; ]_c \ar
FASC:FADD _3 :FLIP   [ \;   FLIP   \; ]_c&k_{3r}\\

r_{35}:&FASC:FADD _3 :CASP8 _2    [ \;   CASP8   \; ]_c
\ar   FASC:FADD _3 :CASP8 _3    [ \;     \; ]_c&k_{3f}\\

r_{36}:&FASC:FADD _3 :CASP8 _3    [ \;     \; ]_c \ar
FASC:FADD _3 :CASP8 _2    [ \;   CASP8   \; ]_c&k_{3r}\\
r_{37}:&FASC:FADD _3 :CASP8 _2    [ \;   FLIP   \; ]_c
\ar   FASC:FADD _3 :CASP8 _2 :FLIP   [ \;     \; ]_c&k_{3f}\\

r_{38}:&FASC:FADD _3 :CASP8 _2 :FLIP   [ \;     \; ]_c
\ar   FASC:FADD _3 :CASP8 _2    [ \;   FLIP   \; ]_c&k_{3r}\\

r_{39}:&FASC:FADD _3 :CASP8:FLIP   [ \;   CASP8   \; ]_c
\ar   FASC:FADD _3 :CASP8 _2 :FLIP   [ \;     \; ]_c&k_{3f}\\
r_{40}:&FASC:FADD _3 :CASP8 _2 :FLIP   [ \;     \; ]_c
\ar   FASC:FADD _3 :CASP8:FLIP   [ \;   CASP8   \; ]_c&k_{3r}\\

r_{41}:&FASC:FADD _3 :CASP8:FLIP   [ \;   FLIP   \; ]_c
\ar
 FASC:FADD _3 :CASP8:FLIP _2    [ \;     \; ]_c&k_{3f}\\

r_{42}:&FASC:FADD _3 :CASP8:FLIP _2    [ \;     \; ]_c
\ar   FASC:FADD _3 :CASP8:FLIP   [ \;   FLIP   \; ]_c&k_{3r}\\

r_{43}:&FASC:FADD _3 :FLIP _2    [ \;   CASP8   \; ]_c
\ar   FASC:FADD _3 :CASP8:FLIP _2    [ \;     \; ]_c&k_{3f}\\

r_{44}:&FASC:FADD _3 :CASP8:FLIP _2    [ \;     \; ]_c
\ar   FASC:FADD _3 :FLIP _2    [ \;   CASP8   \; ]_c&k_{3r}
\end{array}$

\hspace*{-1cm}$
\begin{array}{clc}
\mbox{label} & \ \ \ \ \mbox{ rule }& \mbox{rate}\\

r_{45}:&FASC:FADD _3 :FLIP _2    [ \;   FLIP   \; ]_c \ar
  FASC:FADD _3 :FLIP _3    [ \;     \; ]_c&k_{3f}\\

r_{46}:&FASC:FADD _3 :FLIP _3    [ \;     \; ]_c \ar
FASC:FADD _3 :FLIP _2    [ \;   FLIP   \; ]_c&k_{3r}\\

r_{47}:&FASC:FADD _2    [ \;   CASP8   \; ]_c \ar
FASC:FADD _2 :CASP8   [ \;     \; ]_c&k_{3f}\\

r_{48}:&FASC:FADD _2 :CASP8   [ \;     \; ]_c \ar
FASC:FADD _2    [ \;   CASP8   \; ]_c&k_{3r}\\

r_{49}:&FASC:FADD _2    [ \;   FLIP   \; ]_c \ar
FASC:FADD _2 :FLIP   [ \;     \; ]_c&k_{3f}\\

r_{50}:&FASC:FADD _2 :FLIP   [ \;     \; ]_c \ar
FASC:FADD _2    [ \;   FLIP   \; ]_c&k_{3r}\\

r_{51}:&FASC:FADD _2 :CASP8   [ \;   CASP8   \; ]_c \ar
FASC:FADD _2 :CASP8 _2    [ \;     \; ]_c&k_{3f}\\

r_{52}:&FASC:FADD _2 :CASP8 _2    [ \;     \; ]_c \ar
FASC:FADD _2 :CASP8   [ \;   CASP8   \; ]_c&k_{3r}\\

r_{53}:&FASC:FADD _2 :CASP8   [ \;   FLIP   \; ]_c \ar
FASC:FADD _2 :CASP8:FLIP   [ \;     \; ]_c&k_{3f}\\

r_{54}:&FASC:FADD _2 :CASP8:FLIP   [ \;     \; ]_c \ar
FASC:FADD _2 :CASP8   [ \;   FLIP   \; ]_c&k_{3r}\\

r_{55}:&FASC:FADD _2 :FLIP   [ \;   CASP8   \; ]_c \ar
FASC:FADD _2 :CASP8:FLIP   [ \;     \; ]_c&k_{3f}\\

r_{56}:&FASC:FADD _2 :CASP8:FLIP   [ \;     \; ]_c \ar
FASC:FADD _2 :FLIP   [ \;   CASP8   \; ]_c&k_{3r}\\

r_{57}:&FASC:FADD _2 :FLIP   [ \;   FLIP   \; ]_c \ar
FASC:FADD _2 :FLIP _2    [ \;     \; ]_c&k_{3f}\\

r_{58}:&FASC:FADD _2 :FLIP _2    [ \;     \; ]_c \ar
FASC:FADD _2 :FLIP   [ \;   FLIP   \; ]_c&k_{3r}\\

r_{59}:&FASC:FADD   [ \;   CASP8   \; ]_c \ar
FASC:FADD:CASP8   [ \;     \; ]_c&k_{3f}\\

r_{60}:&FASC:FADD:CASP8   [ \;     \; ]_c \ar   FASC:FADD
  [ \;   CASP8   \; ]_c&k_{3r}\\

r_{61}:&FASC:FADD   [ \;   FLIP   \; ]_c \ar
FASC:FADD:FLIP   [ \;     \; ]_c&k_{3f}\\

r_{62}:&FASC:FADD:FLIP   [ \;     \; ]_c \ar   FASC:FADD
  [ \;   FLIP   \; ]_c&k_{3r}\\

r_{63}:&FASC:FADD _2 :CASP8 _2    [ \;     \; ]_c \ar
FASC:FADD _2    [ \;   CASP8 _2^{P41}    \; ]_c&k_{4}\\

r_{64}:&FASC:FADD _3 :CASP8 _3    [ \;     \; ]_c \ar
FASC:FADD _3 :CASP8   [ \;   CASP8 _2^{P41}    \; ]_c&k_{4}\\

r_{65}:&FASC:FADD _3 :CASP8 _2 :FLIP   [ \;     \; ]_c
\ar   FASC:FADD _3 :FLIP   [ \;   CASP8 _2^{P41}    \; ]_c&k_{4}\\

r_{66}:&FASC:FADD _3 :CASP8 _2    [ \;     \; ]_c \ar
FASC:FADD _3    [ \;   CASP8 _2^{P41}    \; ]_c&k_{4}\\

r_{67}:&  [ \;   CASP8 _2^{P41}    \; ]_c \ar     [ \;
CASP8 _2^*    \; ]_c&k_{5}\\

r_{68}:&  [ \;   CASP8 _2^* , CASP3    \; ]_c \ar     [
\;   CASP8 _2^* :CASP3   \; ]_c&k_{6f}\\

r_{69}:&  [ \;   CASP8 _2^* :CASP3   \; ]_c \ar     [ \;
  CASP8 _2^* , CASP3   \; ]_c&k_{6r}\\

r_{70}:&  [ \;   CASP8 _2^* , CASP3^*   \; ]_c \ar     [ \;
  CASP8 _2^* : CASP3      \; ]_c&k_{7}\\

r_{71}:&  [ \;   CASP8 _2^* , Bid   \; ]_c \ar     [ \;
CASP8 _2^* :Bid   \; ]_c&k_{ 8f }\\

r_{72}:&  [ \;   CASP8 _2^* :Bid   \; ]_c \ar     [ \;
CASP8 _2^* , Bid    \; ]_c&k_{8r}\\

r_{73}:&  [ \;   CASP8 _2^* ,tBid   \; ]_c \ar     [ \;
CASP8 _2^* : Bid   \; ]_c&k_{7}\\

r_{74}:&  [ \;   tBid, Bax   \; ]_c \ar     [ \;
tBid:Bax   \; ]_c&k_{9f}\\

r_{75}:&  [ \;   tBid:Bax   \; ]_c \ar     [ \;   tBid,
Bax   \; ]_c&k_{9r}\\

r_{76}:&  [ \;   tBid:Bax, Bax   \; ]_c \ar     [ \;
tBid:Bax _2     \; ]_c&k_{9f}\\

r_{77}:&  [ \;   tBid:Bax _2     \; ]_c \ar     [ \;
tBid:Bax, Bax    \; ]_c&k_{9r}\\

r_{78}:&tBid:Bax_2   [ \;   Smac   \; ]_m \ar   Smac ^*    [ \;
    \; ]_m&k_{10}\\

r_{79}:&tBid:Bax_2   [ \;   Cyto.c   \; ]_m \ar   Cyto.c ^*    [
\;     \; ]_m&k_{10}\\

r_{80}:&  [ \;   Smac ^* , XIAP   \; ]_c \ar     [ \;
Smac ^* :XIAP   \; ]_c&k_{11f}\\

r_{81}:&  [ \;   Smac ^* :XIAP   \; ]_c \ar     [ \;
Smac ^* , XIAP   \; ]_c&k_{11r}\\

r_{82}:&  [ \;   Cyto.c ^* , Apaf, ATP    \; ]_c \ar     [ \;
  Cyto.c ^* :Apaf:ATP   \; ]_c&k_{12f}\\

r_{83}:&  [ \;   Cyto.c ^* :Apaf:ATP   \; ]_c \ar     [
\;   Cyto.c ^* , Apaf, ATP   \; ]_c&k_{12r}\\

r_{84}:&  [ \;   Cyto.c ^* :Apaf:ATP, CASP9   \; ]_c \ar
    [ \;   Cyto.c ^* :Apaf:ATP:CASP9   \; ]_c&k_{13f}\\

r_{85}:&  [ \;   Cyto.c ^* :Apaf:ATP:CASP9   \; ]_c \ar
  [ \;   Cyto.c ^* :Apaf:ATP, CASP9   \; ]_c&k_{13r}\\

r_{86}:&  [ \;   Cyto.c ^* :Apaf:ATP:CASP9, CASP9  \; ]_c
\ar     [ \;   Cyto.c ^* :Apaf:ATP:CASP9 _2    \; ]_c&k_{14f}\\

r_{87}:&  [ \;   Cyto.c ^* :Apaf:ATP:CASP9 _2    \; ]_c
\ar     [ \;   Cyto.c ^* :Apaf:ATP:CASP9, CASP9   \; ]_c&k_{14r}\\

r_{88}:& [ \;   Cyto.c ^* :Apaf:ATP:CASP9 _2    \; ]_c
\ar     [ \;   Cyto.c ^* :Apaf:ATP:CASP9, CASP9 ^*    \; ]_c&k_{15}\\

r_{89}:&  [ \;   CASP9 ^* , CASP3   \; ]_c \ar     [ \;
CASP9 ^* :CASP3   \; ]_c&k_{16f}\\

r_{90}:&  [ \;   CASP9 ^* :CASP3   \; ]_c \ar     [ \;
CASP9 ^* , CASP3    \; ]_c&k_{16r}
\end{array}$

\hspace*{-1cm}$
\begin{array}{clc}
\mbox{label} & \ \ \ \ \mbox{ rule }& \mbox{rate}\\
r_{91}:&  [ \;   CASP9 ^* :CASP3   \; ]_c \ar     [ \;
CASP9 ^* , CASP3 ^*     \; ]_c&k_{17}\\

r_{92}:&  [ \;   CASP9, XIAP   \; ]_c \ar     [ \;
CASP9:XIAP   \; ]_c&k_{18f}\\

r_{93}:&  [ \;   CASP9:XIAP   \; ]_c \ar     [ \;
CASP9, XIAP   \; ]_c&k_{18r}\\

r_{94}:&  [ \;   CASP3 ^* , XIAP   \; ]_c \ar     [ \;
CASP3 ^* :XIAP   \; ]_c&k_{19f}\\

r_{95}:&  [ \;   CASP3 ^* :XIAP   \; ]_c \ar     [ \;
CASP3 ^* , XIAP   \; ]_c&k_{19r}\\

r_{96}:&Bax   [ \;   Bcl 2    \; ]_m \ar     [ \;
Bcl 2 :Bax   \; ]_m&k_{20f}\\

r_{97}:&  [ \;   Bcl 2 :Bax   \; ]_m \ar   Bax   [ \;
Bcl 2    \; ]_m&k_{20r}\\

r_{98}:&tBid   [ \;   Bcl 2    \; ]_m \ar     [ \;
Bcl 2 :tBid   \; ]_m&k_{20f}\\

r_{99}:&  [ \;   Bcl 2 :tBid   \; ]_m \ar   tBid   [ \;
Bcl 2    \; ]_m&k_{20r}\\

r_{100}:&Vpr  [ \; \; ]_m \ar   Vpr   [ \;
Bcl 2    \; ]_m&k_{21f}\\

r_{101}:&  [ \;   Vpr , Bax   \; ]_c \ar    [ \;
Vpr    \; ]_c&k_{22f}\\

r_{102}:& [ \; Tat \; ]_c \ar  [ \;
 Tat, Casp8\; ]_c&k_{23f}\\

r_{103}:& [ \; Tat \; ]_c \ar  [ \;
 Tat, Bcl2\; ]_c&k_{24f}\\

r_{104}:& [ \; Tat,Bcl2 \; ]_c \ar  [ \;
 Tat\; ]_c&k_{25f}\\

r_{105}:& [ \; HIV Protease,Casp8 \; ]_c \ar  [ \;
 HIV Protease, Casp8*\; ]_m&k_{26f}\\

r_{106}:& HIV Protease [ \; Bcl 2 \; ]_m \ar  HIV Protease [ \;
\; ]_m&k_{27f}\\

r_{107}:& HIVRNA [ \; \; ]_n \ar  [ \;
 HIVRNA \; ]_m&k_{28f}\\

r_{108}:& [ \; HIVRNA \; ]_n \ar  Vpr [ \;
 HIVRNA \; ]_m&k_{29f}\\

r_{109}:& HIVRNA [ \; \; ]_n \ar  Tat [ \;
 HIVRNA \; ]_m&k_{30f}\\

r_{110}:& HIVRNA [ \; \; ]_n \ar  HIV Protease [ \;
 HIVRNA \; ]_m&k_{31f}\\

\end{array}
$
\bigskip

The deterministic kinetic constants (reaction rates) mentioned in the previous table are given in the following;
we refer the interested reader to \cite{FAS} for more details about the rates and references for their estimation.

\begin{eqnarray*}
k_{1f}&=& 9.09E-05\ nM^{-1}s^{-1}\\
k_{1r}&=& 1.00E-04\ s^{-1}\\
k_{2f}&=& 5.00E-04\ nM^{-1}s^{-1}\\
k_{2r}&=& 0.2\ s^{-1}\\
k_{3f}&=& 3.50E-03\ nM^{-1}s^{-1}\\
k_{3r}&=& 0.018\ s^{-1}\\
k_{4}&=& 0.3\ s^{-1}\\
k_{5}&=& 0.1\ s^{-1}\\
k_{6f}&=& 1.00E-05\ nM^{-1}s^{-1}\\
k_{6r}&=& 0.06\ s^{-1}\\
k_{7}&=& 0.1\ s^{-1}\\
k_{8f}&=& 5.00E-03\ nM^{-1}s^{-1}\\
k_{8r}&=& 0.005\ s^{-1}\\
k_{9f}&=& 2.00E-04\ nM^{-1}s^{-1}\\
k_{9r}&=& 0.02\ s^{-1}\\
k_{10}&=& 1.00E-03\ nM^{-1}s^{-1}\\
k_{11f}&=& 7.00E-03\ nM^{-1}s^{-1}\\
k_{11r}&=& 2.21E-03\ s^{-1}\\
k_{12f}&=& 2.78E-07\ nM^{-1}s^{-1}\ nM^{-1}\\
k_{12r}&=& 5.70E-03\ s^{-1}\\
k_{13f}&=& 2.84E-04\ nM^{-1}s^{-1}\\
k_{13r}&=& 0.07493\ s^{-1}\\
k_{14f}&=& 4.41E-04\ nM^{-1}s^{-1}\\
k_{14r}&=& 0.1\ s^{-1}\\
k_{15}&=& 0.7\ s^{-1}\\
k_{16f}&=& 1.96E-05\ nM^{-1}s^{-1}\\
k_{16r}&=& 0.05707\ s^{-1}\\
k_{17}&=& 4.8\ s^{-1}\\
k_{18f}&=& 1.06E-04\ nM^{-1}s^{-1}\\
k_{18r}&=& 1.00E-03\ s^{-1}\\
k_{19f}&=& 2.47E-03\ nM^{-1}s^{-1}\\
k_{19r}&=& 2.40E-03\ s^{-1}\\
k_{20f}&=& 2.00E-03\ nM^{-1}s^{-1}\\
k_{20r}&=& 0.02 s^{-1}\\
k_{21f}&=& 0.0006 s^{-1}\\
k_{22f}&=& 0.000041 nM{-1}s^{-1}\\
k_{23f}&=& 0.0023 s^{-1}\\
k_{24f}&=& 0.00004 s^{-1}\\
k_{25f}&=& 0.00000523 nM^{-1}s^{-1}\\
k_{26f}&=& 0.0000050005 nM^{-1}s^{-1}\\
k_{27f}&=& 0.00000523 nM{-1}s^{-1}\\
k_{28f}&=& 0.000046294 nM{-1}s^{-1}\\
k_{29f}&=& 0.000578703711 nM{-1}s^{-1}\\
k_{30f}&=& 0.000578703709 nM{-1}s^{-1}\\
k_{31f}&=& 0.00057870371 nM{-1}s^{-1}\\
\end{eqnarray*}

\end{document}